# A First-Principles Study on the Adsorption of Small Molecules on Arsenene: Comparison of Oxidation Kinetics in Arsenene, Antimonene, Phosphorene and InSe

Andrey A. Kistanov*[a], Salavat Kh. Khadiullin[b], Sergey V. Dmitriev[a, c] and Elena A. Korznikova[a]

[a] Dr. A.A. Kistanov, Dr. Prof. S.V. Dmitriev, Dr. E.A. Korznikova Institute for Metals Superplasticity Problems, Russian Academy of Sciences 39, Stepana Khalturina st., 450001, Ufa, Russia E-mail: andrei.kistanov.ufa@gmail.com (A. A. K.)

[b] Dr S. Kh. Khadiullin Ufa State Aviation Technical University, 12 Karl Maks st., 450000 Ufa, Russia

[c] Dr. Prof. S.V. Dmitriev National Research Tomsk State University, 36 Prospect Lenina Tomsk, 634050, Russia

**Keywords:** group V two-dimensional materials • arsenene • first-principles calculations • electronic structure • $O_2$ molecule splitting

**Abstract:** Arsenene, a new group V two-dimensional (2D) semiconducting material beyond phosphorene and antimonene, has recently gained an increasing attention owning to its various interesting properties which can be altered or intentionally functionalized by chemical reactions with various molecules. This work provides a systematic study on the interactions of arsenene with the small molecules, including $H_2$, $NH_3$, $O_2$, $H_2O$, $NO$, and $NO_2$. It is predicted that $O_2$, $H_2O$, $NO$, and $NO_2$ are strong acceptors, while $NH_3$ serves as a donor. Importantly, it is shown a negligible charge transfer between $H_2$ and arsenene which is ten times lower than that between $H_2$ and phosphorene and about thousand times lower than that between $H_2$ and InSe and antimonene. The calculated energy barrier for $O_2$ splitting on arsenene is found to be as low as 0.67 eV. Thus, pristine arsenene may easily oxidize in ambient conditions as other group V 2D materials. On the other hand, the acceptor role of $H_2O$ on arsenene, similarly to the cases of antimonene and InSe, may help to prevent the proton transfer between $H_2O$ and $O–$ species by forming acids, which suppresses further structural degradation of arsenene. The structural decomposition of the 2D layers upon interaction with the environment may be avoided due to the acceptor role of $H_2O$ molecules as the study predicts from the comparison of common group V 2D materials. However, the protection for arsenene is still required due to its strong interaction with other small environmental molecules. The present work renders the possible ways to protect arsenene from structure degradation and to modulate its electronic properties, which is useful for the material synthesis, storage and applications.

**Introduction**

In a recent time, several new group V two-dimensional (2D) semiconducting materials, such as phosphorene and antimonene have been predicted by theoretical density functional theory based calculations[1-4] and, later, successfully fabricated experimentally[5-11]. These materials have deserved a great attention of a scientific community due to their phenomenal properties, such as high mechanical stability[2, 12-15], wide and tunable band gap[16-19], asymmetric electronic and phononic transport[20], extraordinary high carrier mobility and on–off current ratio[21].

Very recently, a new group V 2D material, arsenene, has been identified based on theoretical calculations[1, 22-24]. A monolayer arsenene, similarly to phosphorene and antimonene, possesses both puckered and buckled layered structure[22]. Currently, arsenene arouses an intensive attention owing to its intrigue properties such as a wide band gap of ~1.5 eV[22, 23], high electrons and holes mobility[21, 25], low thermal conductivity and anisotropy of both electrical and thermal transport properties[26]. In addition, an indirect to a direct band gap and metal to semiconductor transitions for arsenene have been found under the electric field and strain engineering[22, 27]. All of this suggests arsenene as a promising candidate for various potential applications[28-30], such as sensors, energy storage, and solar cells. Despite the exploration of arsenene is still challenging, it is expected that a monolayer arsenene can be obtained by the mechanical exfoliation or by chemical vapour deposition due to a weak interlayer interaction in bulk arsenic. A significant breakthrough in arsenene fabrication was achieved when the multilayer arsenene nanoribbons with the thickness of 14 nm was synthesized via the plasma-assisted process[31].

Because of its atomically thin structure, large surface area and weak electronic screening, 2D materials, including phosphorene, InSe and antimonene, are easily subjected to the exposure of the environment, external adsorbates and dopants[32-35]. For instance, phosphorene and InSe easily degrade under light illumination and in moisture environment[36-39], while antimonene tends to be more stable due to the lower probability of acids formations under the co-adsorption of $O_2$ and $H_2O$ molecules[40]. Furthermore, external molecules and dopants can remarkably alter the electronic properties and chemical activities of 2D materials through charge carriers transfer or modification of the work function of the material[41, 42].

Selective surface decoration by molecules such as NO, $NO_2$ and $O_2$, and transitional metal atoms has been shown to cause alteration of carrier density, shift of the Fermi level and even change in the optical properties of many 2D materials[43]. For example, the band gap together with the effective mass of the charge carriers of phosphorene are well controlled by changing the concentration of oxygen and fluorine adsorbates[44]. Very recent study has shown that the adsorption of the NO and $NO_2$ gas molecules on arsenene leads to significant modifications of the density of states near the Fermi level[45]. Thereby, more research on the interaction of arsenene with environmental small molecules and common metal dopants is required to enable the fabrication of high quality arsenene on a large scale.

In this work, by using first-principles calculations, the effects of physisorption of typical small molecules, including $H_2$, $NH_3$, $O_2$, $H_2O$, NO, and $NO_2$ on the electronic properties and the charge transfer ability of monolayer arsenene are investigated. Furthermore, the effect of the environmental oxygen and water molecules on the stability of arsenene is considered from the atomic scale. In addition, the adsorption and oxidation ability of arsenene and its counterparts, such as phosphorene, InSe and antimonene are compared. The present study provides new knowledge which is of a critical importance for the synthesis, storage, and applications of arsenene.

**Results and Discussion**

Main Text Paragraph. The adsorption of small molecules, such as $H_2$, $NH_3$, $O_2$, $H_2O$, NO, and $NO_2$ on the arsenene surface is systematically studied. Particularly, the electronic properties and chemical activities of molecule−adsorbed arsenene and the adsorption energy and charge transfer between these small molecules and the arsenene surface are considered.

For each molecule, several different configurations and possible adsorbing sites are considered, including the site atop the centre of the hexagon, atop the As site, and the site atop the As-As bond with the molecules being aligned parallel, perpendicular or tilted to the surface (for more details see Supporting Information). The results of the adsorption energy $E_a$, the charge transfer $\Delta q$ and the shortest distance d from the molecule to the arsenene surface for the lowest-energy configuration are presented in Table 1.

*$H_2$ adsorption.* Figure 1a shows the lowest energy configuration for the $H_2$ molecule adsorbed on arsenene. The molecule locates directly above the As atom with the H−H bond (bond length is 0.75 Å) being aligned perpendicular to the arsenene surface at distance d of 2.84 Å. The adsorption energy $E_a$ of the $H_2$ molecule on arsenene is -0.05 eV which is about 2 times higher than that of phosphorene, while it is comparable with that of graphene[41], antimonene and InSe [34]. Given the fact that graphene is one of the most popular materials for a hydrogen storage due to its ability for stable hydrogen storage together with the effortless hydrogen release[46], the predicted $H_2$ adsorption energy for arsenene suggests this material as a promising competitor to graphene in use for hydrogen storage devices.

The local density of states (LDOS) (Figure 1b) and band structure (Figure 1c) analyses reveal that there are no additional $H_2$-induced states in the vicinity of the arsenene band gap. As a result, the band gap size of arsenene adsorbed with the $H_2$ molecule is the same as that of pristine arsenene (1.36 eV). This picture is typical for most 2D counterparts of arsenene, such as phosphorene, InSe and antimonene[33, 34, 39]. Owing to the similar presence of lone-pair electrons of the surface of arsenene and phosphorene, InSe and antimonene, it is clearly needs to compare the charge doping behaviour of $H_2$ among these materials. The Bader charge transfer analysis reveals a negligibly small charge transfer $\Delta q$ of 0.001 e per molecule from arsenene to the $H_2$ molecule. Interestingly, the acceptor role of the $H_2$ molecule makes arsenene different from phosphorene, InSe and antimonene where the $H_2$ molecule serves as a strong donor.

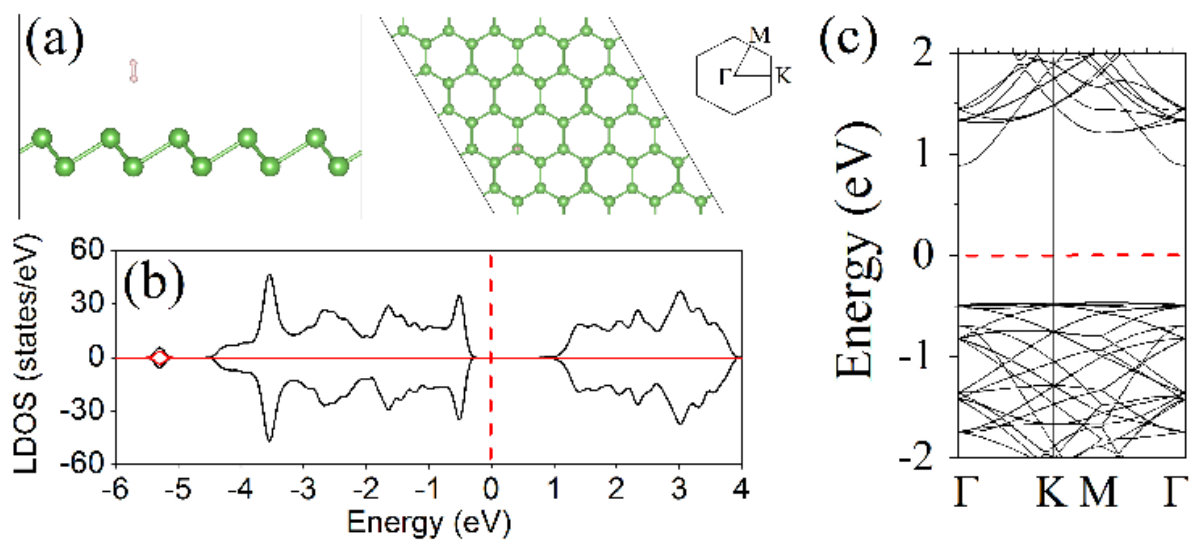

**Figure 1.** (a) The top and side views of the lowest-energy configuration for arsenene adsorbed with the $H_2$ molecule. (b) The band structure of arsenene adsorbed with the $H_2$ molecule. (c) The total DOS (black line) and LDOS (red line) of arsenene adsorbed with the $H_2$ molecule. The red dashed lines show the Fermi level.

**NH$_3$ adsorption.** The lowest energy configuration for the NH$_3$ molecule adsorbed on arsenene is presented in Figure 2a. The molecule is located at d = 2.55 Å at the hollow hexagon centre with the N atom pointing towards the surface and the three H atoms pointing away from the surface. The calculated $E_a$ of -0.25 eV suggests the strongest interaction of the NH$_3$ molecule with arsenene compared to phosphorene, InSe and antimonene (see Table 1).

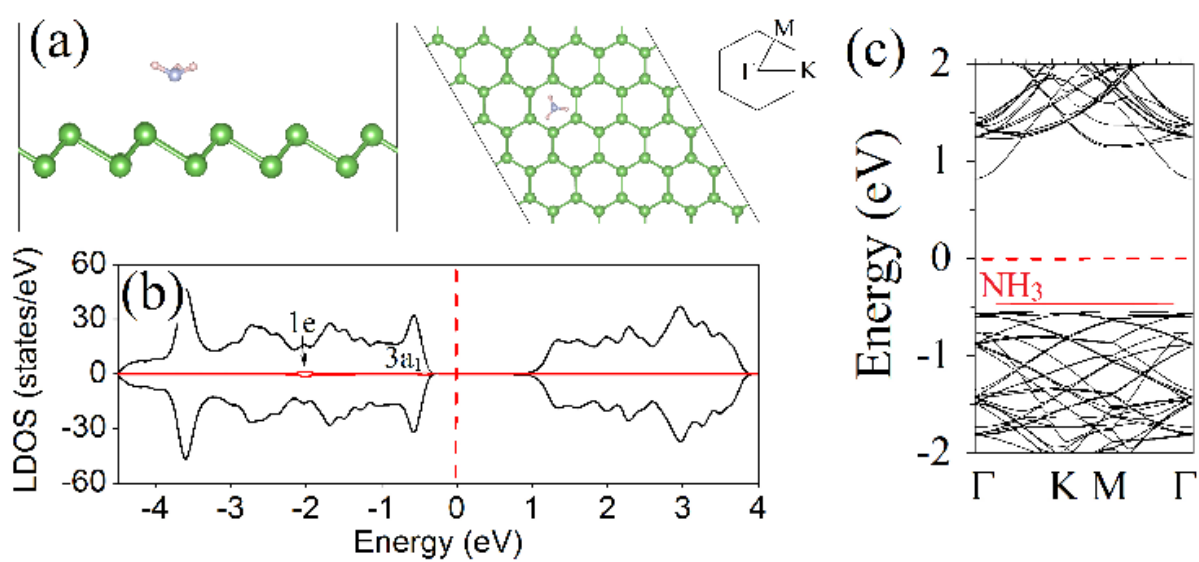

**Figure 2.** The same as in Figure 1 but for arsenene adsorbed with the NH$_3$ molecule. The bands coloured in black and red represent arsenene and NH$_3$, respectively.

Figure 2b shows the LDOS for the NH$_3$ molecule adsorbed on arsenene. It is seen that the doubly degenerated 1e state (HOMO) is significantly below the Fermi level and more affected by the As atoms, which is evidenced by the strongly broadened LDOS. The nonbonding 3a$_1$ states appear in the vicinity of the valance band of arsenene, which is proved by the band structure analysis (Figure 2c). In addition, the charge transfer analysis predicts the NH$_3$ molecule as a donor to arsenene. The total amount of the charge transferred from the molecule to the arsenene surface is 0.022 e. A similar donor behaviour is predicted for the NH$_3$ molecule adsorbed on graphene[41] and phosphorene[33, 36], while for InSe[34] and antimonene the NH$_3$ molecule behaves as an acceptor.

Table 1. The adsorption energy $E_a$, the amount of charge transfer $\Delta q$, the shortest distance d from the molecule to the surface, and the donor/acceptor characteristics of the molecular dopant on the surface. Note that a positive (negative) $\Delta q$ indicates a loss (gain) of electrons from each molecule to the surface.

| Molecule | Arsenene | | | | Phosphorene [Ref. 33] | | InSe [Ref. 34] | | Antimonene [Ref. 40] | |
|---|---|---|---|---|---|---|---|---|---|---|
| | d [Å] | Doping nature | $E_a$ [eV] | $\Delta q$ [e] | $E_a$ [eV] | $\Delta q$ [e] | $E_a$ [eV] | $\Delta q$ [e] | $E_a$ [eV] | $\Delta q$ [e] |
| H$_2$ | 2.84 | acceptor | −0.05 | −0.001 | −0.13 | 0.013 | −0.05 | 0.146 | −0.04 | 0.138 |
| NH$_3$ | 2.55 | donor | −0.25 | 0.022 | −0.18 | 0.050 | −0.20 | −0.019 | −0.12 | −0.029 |
| H$_2$O | 2.61 | acceptor | −0.19 | −0.016 | −0.14 | 0.035 | −0.17 | −0.01 | −0.20 | −0.021 |
| O$_2$ | 2.64 | acceptor | −0.54 | −0.076 | −0.27 | −0.064 | −0.12 | −0.001 | −0.61 | −0.116 |
| NO | 2.23 | acceptor | −0.27 | −0.158 | −0.32 | −0.074 | −0.13 | −0.094 | −0.44 | −0.067 |
| NO$_2$ | 2.51 | acceptor | −0.47 | −0.344 | −0.50 | −0.185 | −0.24 | −0.039 | −0.81 | −0.156 |

**H$_2$O and O$_2$ adsorptions.** Further, the interaction of the H$_2$O and O$_2$ molecules which commonly affect the structural stability of 2D materials, with arsenene surface is considered. The most energetically stable configuration for the H$_2$O and O$_2$ molecules adsorbed on arsenene are presented in Figures 3a and d, respectively. The H$_2$O molecule adopts almost flat alignment with the two O−H bonds only slightly tilted to the arsenene surface and is located at the centre of the hexagon at $d = {\sim}2.61$ Å. The H$_2$O molecule possesses a relatively weak $E_a$ = -0.19 eV which is similar to phosphorene, InSe and antimonene.

The O$_2$ molecule is slightly tilted to the arsenene surface and is located above the As atom at $d$ = 3.64 Å. Surprisingly, the O$_2$ molecule has a high $E_a$ = -0.54 eV on arsenene which is comparable with that for antimonene and larger than that for phosphorene and InSe. Clearly, O$_2$ has a strong physisorption on arsenene, and the O−O bond increases from 1.22 Å (the free gas molecule) to 1.25 Å (the bound molecule). Figures 3b and c, where the LDOS and band structure plots for arsenene adsorbed with the H$_2$O molecule are presented, indicate the absence of H$_2$O-induced states within the band gap of arsenene. It should be noted that the 1b$_2$ and 1b$_1$ orbitals of the H$_2$O molecule are significantly broadened and coincide with the valence states of arsenene (Figure 3b), which suggests that the durability and carrier mobility of arsenene may be affected by the presence of water due to the strong state coupling.

The LDOS (Figure 3e) and band structure (Figure 3f) of arsenene adsorbed with the O$_2$ molecule reflect additional O$_2$-induced states within the band gap of arsenene. Despite a slight broadening of the 2π HOMO state (Figure 3e) the orbitals of O$_2$ are less affected by arsenene. The antibonding 2π$^*$ LUMO is nearly unaffected and is located at 0.44 eV above the Fermi level (Figure 3f) within the band gap of arsenene. In addition, the band structure analysis predicts a negligible decrease of the band gap size of arsenene upon the $O_2$ molecule adsorption from 1.36 (pristine arsenene) to 1.35 eV (molecule-adsorbed arsenene). Thereby, the presence of the $O_2$-induced states within the original band gap of arsenene together with strong adsorption and oxidation abilities of the $O_2$ molecule to arsenene can significantly influence the electronic properties of arsenene. The charge transfer analysis shows that both the $H_2O$ and $O_2$ molecule are acceptors to arsenene with a total charge transfer of 0.016 and 0.076 e per molecule, respectively.

Therefore, arsenene demonstrates a weaker interaction with the $H_2O$ molecule, like graphene[41,46], phosphorene[33] and antimonene[38]. Nevertheless, arsenene possesses a high oxidation ability and may oxidize easily at ambient conditions. The oxidation of arsenene, similarly to phosphorene and antimonene, originates from the presence of $O_2$ rather than $H_2O$, due to a stronger binding strength and a larger amount of charge transfer of $O_2$.

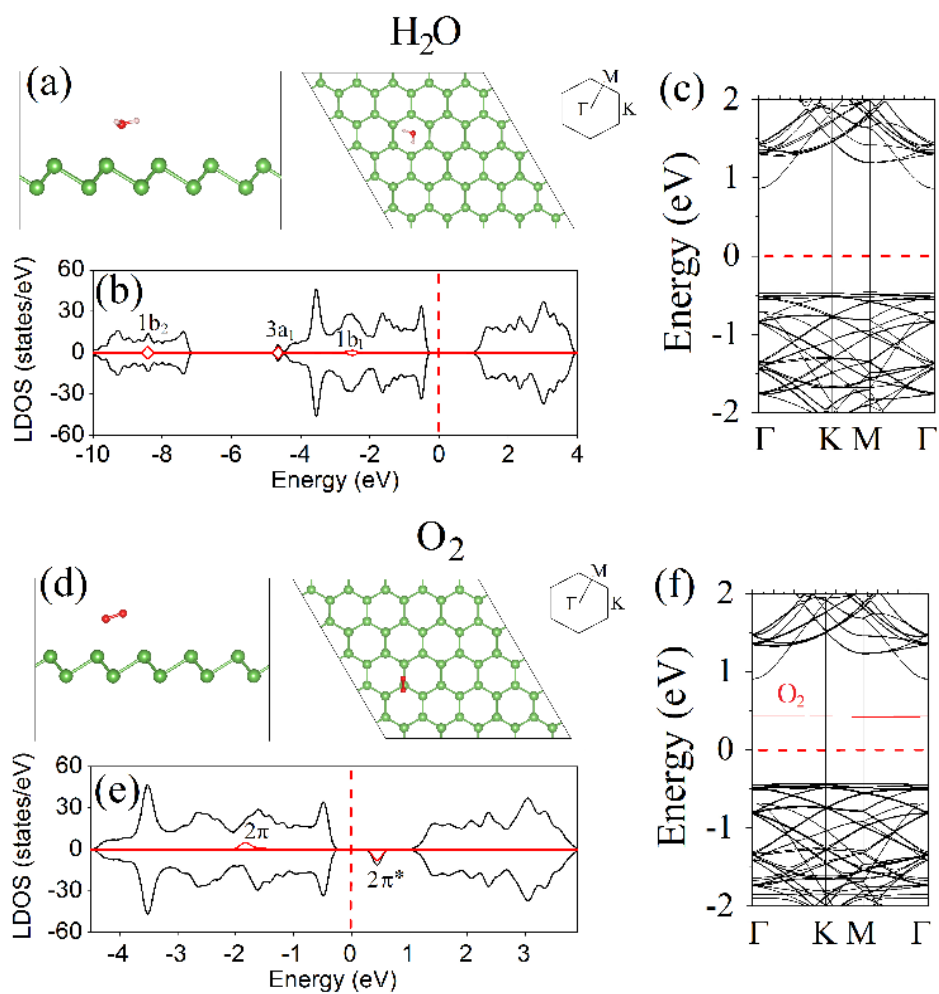

**Figure 3.** The same as in Figure 1 but for arsenene adsorbed with the $H_2O$ and $O_2$ molecules. The bands coloured in black and red represent arsenene and $O_2$, respectively.

**NO adsorption.** Figure 4a presents the most energetically favourable configuration for the NO molecule adsorbed on arsenene. NO adopts a tilted configuration and is located above the As-As bond with d = 2.23 Å. As a typical open-shell molecule, NO has a strong interaction with arsenene, which is evidenced by the comparably low $E_a$ = -0.27 eV. Interestingly, the NO molecule on arsenene possesses a moderate $E_a$ which is slightly lower than that for phosphorene (-0.32 eV)[33] and antimonene (-0.44 eV)[39], while it is larger than that for InSe (-0.13 eV)[34].

The LDOS analysis predicts the state hybridization and a strong charge transfer between the NO molecule and arsenene, which induce broadening and splitting of the degeneracy of NO orbitals. Particularly, the degeneracy of the 2π orbital is lifted and split into two levels located close to the conduction band minimum and coincides with the conduction bands of arsenene (see Figure 4b). Moreover, the spin-splitting of the NO molecule level induces a magnetic moment of 1 µB in the adsorbed system. In addition, the band gap of arsenene adsorbed with the NO molecule slightly increases from 1.36 eV (pristine arsenene) to 1.38 eV (molecule-adsorbed arsenene). The band structure plot (Figure 4c) also shows the presence of the NO-induced states within the band gap of arsenene, which may lead to significant modifications of the optical and electronic properties of arsenene due to the possible role of the NO molecule as an electron trapping centre. The charge transfer analysis shows a similar trend of charge transfer for the NO molecule (acceptor) adsorbed on arsenene, phosphorene[32, 33], InSe[34], and antimonene[39]. The total amount of the transferred charge from the arsenene surface to the NO molecule is 0.067 e.

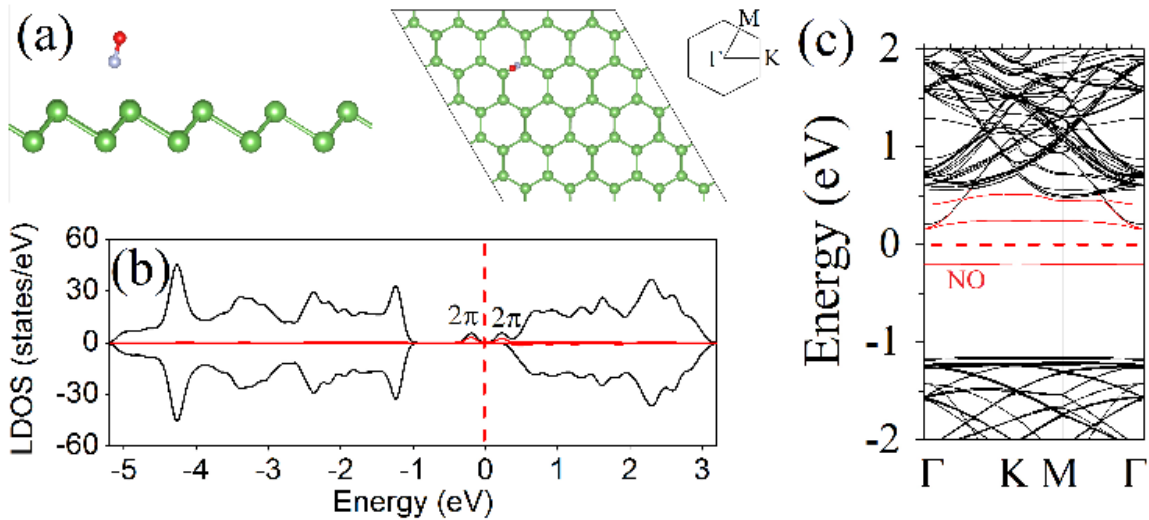

**Figure 4.** The same as in Figure 1 but for arsenene adsorbed with the NO molecule. The bands coloured in black and red represent arsenene and NO, respectively.

**NO₂ adsorption.** It has been predicted both theoretically and experimentally[48, 49] that among all typical small molecules the NO₂ molecule possesses the strongest interaction with phosphorene[33] and antimonene[39]. Here, for the NO₂ molecule adsorbed on arsenene, the $E_a$ = -0.47 eV which is much larger than that of InSe ($E_a$ = -0.24 eV) and is comparable to that of phosphorene ($E_a$ = -0.50 eV).

Figure 5a shows the most stable configuration for the NO₂ molecule adsorbed on arsenene. The molecule takes the position above the As-As bond closer to the As atom with the two O atoms situated closer to the arsenene surface with d = 2.51 Å. The strong interaction of the NO₂ molecule with the arsenene surface leads to the increase of the N−O bond length from the 1.20 Å (free NO₂ gas molecule) up to 1.24 Å (the bonded molecule).

The LDOS analysis (Figure 5b) shows that the $6a_1$ orbital is split into two levels and is located within the band gap of arsenene. The LUMO ($6a_1$, spin-down) and the HOMO ($6a_1$, spin-up) states are located just above and below the Fermi level, respectively, which leads to a magnetic moment of 1 μB. Furthermore, the $4b_1$, $1a_2$ (coincided with the valence states of arsenene) and $5b_1$ (coincided with the conduction states of arsenene) orbitals of the NO₂ molecules are significantly broadened. Such orbital mixing and hybridization result in the enhanced charge transfer between the NO₂ molecule and arsenene surface.

The Bader analysis also suggests a large electron transfer of 0.344 e per molecule from the arsenene surface to the NO₂ molecule. The band structure plot in Figure 5c shows the presence of the NO₂-induced localized states within the band gap of arsenene, which suggests remarkable changes in the optical and electronic properties of arsenene upon adsorption of the NO₂ molecule.

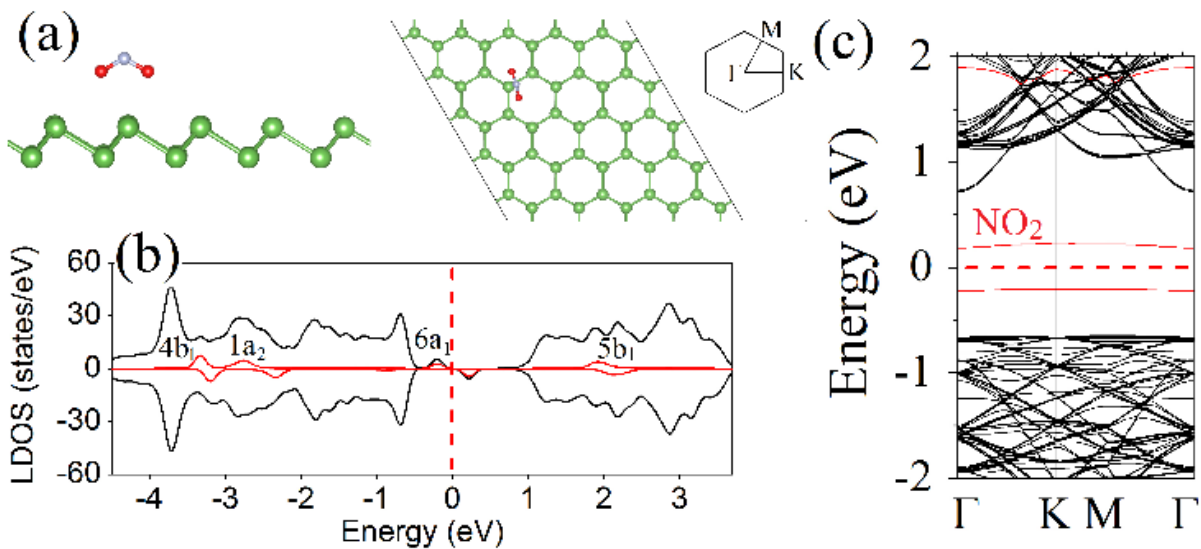

**Figure 5.** The same as in Figure 1 but for arsenene adsorbed with the NO₂ molecule. The bands coloured in black and red represent arsenene and NO₂, respectively.

**Comparison of oxidation kinetics in arsenene, antimonene, phosphorene and InSe.** The $O_2$ molecules significantly affect the stability and performance of 2D materials by means of their oxidation. As it has been mentioned above, the binding energy between the $O_2$ molecule and arsenene (-0.12 eV) is about two times lower than that between the $O_2$ molecule and phosphorene (-0.27 eV) and about five times lower than that between the $O_2$ molecule and antimonene (-0.61 eV). Similarly, the charge transfer between the $O_2$ molecule and arsenene (-0.076 e) is slightly larger than that between the $O_2$ molecule and phosphorene (-0.064 e) but significantly lower than that between the $O_2$ molecule and antimonene (-0.116 e). Such charge transfer behaviour of arsenene is reasonable since As is only slightly less electronegative than P but more electronegative than Sb.

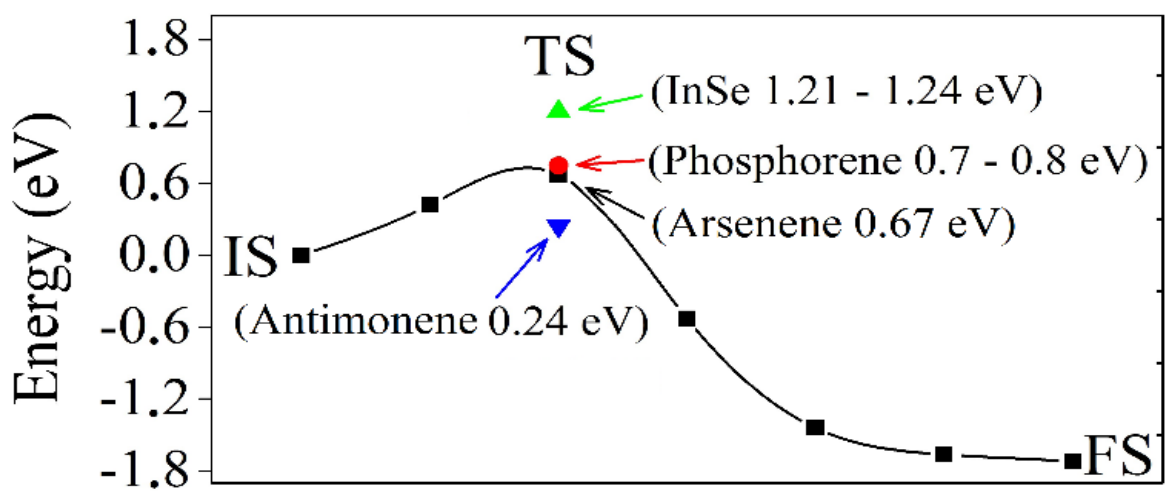

**Figure 6.** The activation barrier for the splitting of the $O_2$ molecule on arsenene (black line), phosphorene (red dot), InSe (green triangle) and antimonene (blue triangle). The energy profile is obtained by the NEB calculation. The the activation barriers of $O_2$ splitting on phosphorene, InSe and antimonene are adopted from Refs. 36, 40, 50-52.

To consider further the oxidization behaviour of arsenene the kinetic analysis on the $O_2$ molecule splitting on its surface is conducted. In addition, the comparison of the oxidization behaviour of arsenene with its counterparts which are phosphorene, InSe, and antimonene is presented. Figure 6 shows the activation barriers for the splitting of the $O_2$ molecule on arsenene, phosphorene, InSe and antimonene. The energy barrier for the decomposition of $O_2$ molecule into two apical –O groups is found as low as ~0.67 eV. Such a small barrier implies that arsenene may easily undergo oxidation in ambient conditions.

Our results well complement recent predictions of a poor stability of arsenene on contact with the air[50]. However, unlike phosphorene which tends to degrade as a result of the formation of acids due to the reaction of oxygen species with the environmental $H_2O$ molecules[53], the cooperative effect of the $O_2$ and $H_2O$ molecules on the stability of arsenene is expected to be different. As it has been reported previously[33,

34, 39] the $O_2$ molecule plays the acceptor role in phosphorene and antimonene, while the $H_2O$ molecule is a donor for phosphorene and an acceptor for antimonene. Hence, the acceptor role of the $H_2O$ molecule may result in the electrostatic repulsion between $[H_2O]^{-\delta}$ and –O group (also negatively charged $-O^{-\gamma}$ with γ being a small positive real number) forming on the antimonene surface, and it leads to a comparably high stability of antimonene in a moisture environment. Our theory is also supported by the recent theoretical investigations which reveal that the water dissociation is significantly promoted on two-dimensional surfaces containing the pre-adsorbed $O_2$.[54-56]

Thereby, based on the kinship of arsenene and antimonene in terms of an acceptor role of the $H_2O$ molecule, it can be proposed that arsenene possesses a similarly low affinity to $H_2O$ molecules as antimonene. Moreover, our results are also consistent with the very recent findings which have predicted that the $H_2O$ molecules remain intact while the $O_2$ molecules are dissociated spontaneously upon interaction with the arsenene surface[57]. Consequently, the comparison of the $H_2O$ molecules charge transfer behaviour in InSe, antimonene and arsenene clearly shows that systems with water molecules acting as acceptors tend to be more stable as they are less likely to form acids under the co-adsorption of $O_2$ and $H_2O$ molecules [39, 51].

**Conclusions**

The first−principles calculations are performed to investigate the adsorption of small molecules, such as $H_2$, $NH_3$, $O_2$, $H_2O$, NO, and $NO_2$ on arsenene. It is found that $O_2$, $H_2O$, NO, and $NO_2$ are acceptors to arsenene, while $NH_3$ serves as a donor. Interestingly, the charge transferred between $H_2$ and arsenene is significantly lower than that between $H_2$ and phosphorene, InSe and antimonene. It is also found that $O_2$ serves as a strong acceptor to arsenene, similarly to the case of antimonene. Moreover, the examination of the $O_2$ molecule splitting on arsenene shows a low barrier of 0.67 eV, which suggests that the performance of arsenene tends to be highly sensitive to the environmental $O_2$ molecule. However, the stable surface oxidation layer may be helpful for protecting the underneath arsenene layer against its degradation upon interaction with environmental molecules.

**Computational details**

The density functional theory-based calculations were performed by using Vienna *ab initio* simulation package (VASP)[58]. The van der Waals-corrected functional with Becke88 optimization (optB88) was adopted to consider noncovalent chemical interactions between arsenene and small molecules and atoms[59]. A 5×5×1 arsenene supercell (50 As atoms) was created for adsorption of small molecules and atoms. All the considered structures were fully relaxed until atomic forces and total energy were smaller than 0.01 eV/Å and $10^{-6}$ eV, respectively. Periodic boundary conditions were applied in the in-plane directions, whereas a vacuum separation distance of 20 Å were implemented to eliminate spurious interactions between replicate units. The first Brillouin zone was sampled with a 5×5×1 k-mesh grid and a kinetic energy cutoff of 400 eV were adopted. The Perdew–Burke–Ernzerhof (PBE) functional under the generalized gradient approximation (GGA)[60] gives the relaxed lattice constants of monolayer arsenene of $a = b = 3.76$ Å and the band gap of 1.36 eV, which is well consistent with recent works[22, 27, 61]. The adsorption energy $E_a$ of a molecule (atom) on arsenene was calculated as

$$E_a = E_{As+mol} - E_{As} - E_{mol} \qquad (1)$$

where $E_{As+mol}$, $E_{As}$, and $E_{mol}$ are the energies of arsenene adsorbed with a molecule, the isolated arsenene, and the isolated molecule, respectively. The Bader analysis[62] was applied to estimate the charge transfer between the arsenene surface and small molecules and atoms. The reaction barriers were calculated by using the climbing image nudged elastic band (NEB) method[63].

[1]. S. Zhang, Z. Yan, Y. Li, Z. Chen, H. Zeng, *Angew. Chem. Int. Ed.* **2015**, *54*, 3112−3115.

[2]. G. Wang, R. Pandey, S. P. Karna, *ACS Appl. Mater. Inter.* **2015**, *7*, 11490–11496.

[3]. Z. Zhu, D. Tománek, *Phys. Rev. Lett.* **2014**, *112*, 176802.

[4]. O. Ü. Aktürk, V. O. Özçelik, S. Ciraci, *Phys. Rev. B: Condens. Matter. Mater. Phys.* **2015**, *91*, 235446.

[5]. D. Hanlon, et al., *Nat. Commun.* **2015**, *6*, 8563.

[6]. P. Yasaei, B. Kumar, T. Foroozan, C. Wang, M. Asadi, D. Tuschel, J. E. Indacochea, R. F. Klie, A. S. Khojin, *Adv. Mater.* **2015**, *27*, 1887−1892.

[7]. A. Castellanos-Gomez, et al., *2D Mater.*, **2014**, *1*, 025001.

[8]. P. Ares, F. Aguilar−Galindo, D. Rodríguez−San−Miguel, D. A. Aldave, S. Díaz−Tendero, M. Alcamí, F. Martín, J. Gomez−Herrero, F. Zamora, *Adv. Mater.* **2016**, *28*, 6332−6336.

[9]. J. Kang, S. A. Wells, J. D. Wood, J. H. Lee, X. Liu, C. R. Ryder, J. Zhu, J. R. Guest, C. A. Husko, M. C. Hersam, *Proc. Natl. Acad. Sci.* **2016**, *113*, 11688–11693.

[10]. J. Ji et al., *Nat. Commun.* **2016**, *7*, 13352.

[11]. C. Gibaja et al., *Angew. Chem. Int. Ed.* **2016**, *55*, 14345.

[12]. J. W. Jiang, H. S. Park *J. Phys. D: Appl. Phys.* **2014**, *47*, 385304.

[13]. Q. Wei, X. Peng, *Appl. Phys. Lett.* **2014**, *104*, 251915.

[14]. G. Yang, T. Ma, X. Peng, *Appl. Phys. Lett.* **2018**, *112*, 241904.

[15]. A. A. Kistanov, Y. Cai, K. Zhou, S. V. Dmitriev, Y. W. Zhang, *J. Phys. Chem. C* **2016**, *120*, 6876–6884.

[16]. J. Guan, Z. Zhu, D. Tománek, *Phys. Rev. Lett.* **2014**, *113*, 046804.

[17]. A. S. Rodin, A. Carvalho, A. H. Castro Neto, *Phys. Rev. B* **2014**, *90*, 075429.

[18]. P. Ares, J. J. Palacios, G. Abellán, J. Gómez-Herrero, F. Zamora, *Adv. Mater.* **2018**, *30*, 1703771.

[19]. S. Zhang, M. Xie, F. Li, Z. Yan, Y. Li, E. Kan, W. Liu, Z. Chen, H. Zeng *Angew. Chem. Int. Ed.* **2016**, *55*, 1666–1669.

[20]. Y. Cai, Q. Ke, G. Zhang, Y. P. Feng, V. B. Shenoy, Y. W. Zhang, *Adv. Funct. Mater*. **2015**, *25*, 2230−2236.

[21]. J. Zhao, C. Liu, W. Guo, J. Ma, *Nanoscale* **2017**, *9*, 7006–7011.

[22]. C. Kamal, M. Ezawa, *Phys. Rev. B* **2015**, *91*, 085423.

[23]. L. Kou, Y. Ma, X. Tan, T. Frauenheim, A. Du, S. Smith, *J. Phys. Chem. C* **2015**, *119*, 6918–6922.

[24]. Z. Zhu, J. Guan, D. Tománek, *Phys. Rev. B* **2015**, *91*, 161404.

[25]. Z. Zhang, J. Xie, D. Yang, Y. Wang, M. Si, D. Xue, *Appl. Phys. Express*. **2015**, *8*, 055201.

[26]. C. Wang, Q. Xia, Y. Nie, M. Rahman, G. Guo, *AIP Advances* **2016**, *6*, 035204.

[27]. D. Kecik, E. Durgun, S. Ciraci, *Phys. Rev. B* **2016**, *94*, 205409.

[28]. Y. J. Wang, K. G. Zhou, G. Yu, X. Zhong, H. L. Zhang, *Sci. Rep.* **2016**, *6*, 24981.

[29]. Y. Wang et al., *ACS Appl Mater Interfaces* **2017**, *9*, 29273–29284.

[30]. Y. Wang, Y. Ding *Nanoscale Res. Lett.* **2015**, *10*, 254.

[31]. H. S. Tsai, S. W. Wang, C. H. Hsiao, C. H. Chen, H. Ouyang, Y. L. Chueh, H. C. Kuo, J. H. Liang, *Chem. Mater.* **2016**, *28*, 425–429.

[32]. L. Kou, T. Frauenheim, C. Chen, *J. Phys. Chem. Lett.* **2014**, *5*, 2675–2681.

[33]. Y. Cai, Q. Ke, G. Zhang, Y. W. Zhang, *J. Phys. Chem. C* **2015**, *119*, 3102–3110.

[34]. Y. Cai, G. Zhang, Y. W. Zhang, *J. Phys. Chem. C* **2017**, *121*, 10182–10193.

[35]. Sh. Ma, D. Yuan, Y. Wang, Zhaoyong Jiao, *J. Mater. Chem. C*, **2018**, *6*, 8082–8091.

[36]. A. A. Kistanov, Y. Cai, K. Zhou, S. V. Dmitriev, Y. W. Zhang, *2D Mater.* **2017**, *4*, 015010.

[37]. J. Gao, G. Zhang, Y. W. Zhang, *Nanoscale* **2017**, *9*, 4219−4226.

[38]. N. Liu, S. Zhou, *Nanotechnology* **2017**, *28*, 175708.

[39]. A. A. Kistanov, Y. Cai, D. R. Kripalani, K. Zhou, S. V. Dmitriev, Y. W. Zhang, *J Mater Chem C* **2018**, *6*, 4308−4317.

[40]. A. C. Crowther, A. Ghassaei, N. Jung, L. E. Brus, *ACS Nano* **2012**, *6*, 1865–1875.

[41]. O. Leenaerts, B. Partoens, F. M. Peeters, *Phys. Rev. B: Condens. Matter. Mater. Phys.* **2008**, 77, 125416.

[42]. Z. Chen et al., *ACS Nano* **2014**, *8*, 2943–2950.

[43]. J. Xiao, M. Long, X. Li, Q. Zhang, H. Xu, K. S. Chan, *J. Phys.: Condens. Matter.* **2014**, *26*, 405302.

[44]. S. Nahas, B. Ghosh, S. Bhowmick, A. Agarwal, *Phys. Rev. B* **2016**, *93*, 165413.

[45]. O. Ü. Aktürk, E. Aktürk, S. Ciraci, *Phys. Rev. B*, **2016**, *93*, 035450.

[46]. C. Ataca, E. Aktürk, S. Ciraci, H. Ustunel, *Appl. Phys. Lett.* **2008**, *93*, 043123.

[47]. A. A. Kistanov, Y. Cai, Y. W. Zhang, S. V. Dmitriev, K. Zhou, *J. Phys.: Condens. Matter.* **2017**, *29*, 095302.

[48]. S. Cui, H. Pu, S. A. Wells, Z. Wen, S. Mao, J. Chang, M. C. Hersam, J. Chen, *Nat. Commun.*, **2015**, *6*, 8632.

[49]. A. N. Abbas, B. Liu, L. Chen, Y. Ma, S. Cong, N. Aroonyadet, M. Köpf, T. Nilges, C. Zhou, *ACS Nano* **2015**, *9*, 5618–5624.

[50]. Y. Guo, S. Zhou, Y. Bai, J. Zhao, *ACS Appl Mater Interfaces* **2017**, *9*, 12013−12020.

[51]. A. A. Kistanov, Y. Cai, K. Zhou, S. V. Dmitriev, Y. W. Zhang, *J. Mater. Chem. C* **2018**, *6*, 518–525.

[52]. D. Ma, T. Li, D. Yuan, C. He, Z. Lu, Z. Lu, Z. Yang, Y. Wang, *Appl. Surf. Sci.* **2018**, *434*, 215–227.

[53]. Y. Huang et al., *Chem. Mater.* **2016**, *28*, 8330–8339.

[54]. M. C. Valero, P. Raybaud, J. Phys. Chem. C, 2015, 119, 23515–23526.

[55]. F. F. Ma, S. H. Ma, Z.Y. Jiao, X. Q. Dai, Appl. Surf. Sci., 2016, 384, 10–17.

[56]. G. Wang, W. J. Slough, R. Pandey, S. P. Karna, 2D Mater., 2016, 3, 025011.

[57]. F. Ersan, E. Aktürk, S. Ciraci. *J. Phys. Chem. C* **2016**, *120*, 14345−14355.

[58]. G. Kresse, J. Furthmuller, *Phys. Rev. B: Condens. Matter. Mater. Phys.* **1996**, *54*,11169.

[59]. A. D. Becke, *Phys. Rev. A: At. Mol. Opt. Phys.* **1988**, *38*, 3098.

[60]. J. P. Perdew, K. Burke, M. Ernzerhof, *Phys. Rev. Lett.* **1996**, *77*, 3865−3868.

[61]. M. Y. Liu, Y. Huang, Q. Y. Chen, C. Cao, Y. He, *Sci. Rep.* **2016**, *6*, 29114.

[62]. R. F. W. Bader, *Atoms in Molecules - A Quantum Theory*, Oxford University Press: New York, **1990**.

[63]. G. Henkelman, B. P. Uberuaga, H. Jonsson, *J. Chem. Phys.* **2000**, *113*, 9901.